\documentclass[11pt]{article}

\usepackage{fullpage}
\usepackage{url}
\usepackage{latexsym}
\usepackage{theorem}
\usepackage{amsmath}
\usepackage{verbatim}

\newtheorem{theorem}{Theorem}
\newtheorem{claim}[theorem]{Claim}
\newtheorem{lemma}[theorem]{Lemma}

\newtheorem{definition}[theorem]{Definition}

\newcommand{\val}{{\bf value}}
\newcommand{\Xomit}[1]{}
\newcommand{\cl}{{\bf clicks}}
\newcommand{\cpc}{{\bf cpc}}
\newcommand{\ctr}{{\bf ctr}}
\newcommand{\cost}{{\bf cost}}
\newcommand{\sol}{\mbox{\boldmath$b$}}
\newcommand{\solv}{{\mbox{\boldmath$b$}_V}}
\newcommand{\solu}{{\mbox{\boldmath$b$}_U}}
\newcommand{\soln}{{\mbox{\boldmath$b$}_N}}

\newenvironment{proof}[1][Proof. ]{{\bf #1}}{\qed} 
\newenvironment{proofof}[1]{\noindent {\bf{Proof of #1.}}}{\qed}

\newcommand{\qed}{\hspace*{\fill}$\Box$\\}

\begin{document}

\title{Stochastic Models for Budget Optimization in Search-Based Advertising}

\author{
S. Muthukrishnan\thanks{Google, Inc., New York, NY.} 
\and Martin P\'al\thanks{Google, Inc., New York, NY.}
\and Zoya Svitkina\thanks{Department of Computer Science, Dartmouth College. This work was done while visiting Google, Inc., New York, NY.}
}


\maketitle

\begin{abstract}
Internet search companies sell advertisement slots based on users' search 
queries via an auction. Advertisers have to determine
how to place bids on the keywords of their interest in order to 
maximize their return for a given  
budget: this is the {\em budget optimization} problem. The solution
depends on the distribution of future queries. 
In this paper, we formulate {\em stochastic} versions of 
the budget optimization problem based on natural probabilistic models of 
distribution over future queries, and address  two questions that arise. 
  \begin{description}
    \item[Evaluation] Given a solution, can we evaluate the expected 
    value of the objective function?
    \item[Optimization] Can we find a solution that maximizes 
    the objective function in expectation? 
  \end{description}
Our main results are approximation and complexity results for these two problems
in our three stochastic models.
In particular, our algorithmic results show
that simple {\em prefix} strategies that bid on all cheap keywords 
up to some level are either optimal or good approximations for many cases; 
we show other cases to be NP-hard. 
\end{abstract}

\section{Introduction}

Internet search companies use auctions to sell advertising slots in response to
users' search queries.  To participate in these auctions, an advertiser selects a set of keywords that are 
relevant or descriptive of her business, and submits a bid for each of them. 
Upon seeing a user's query, the search company runs an auction among the advertisers who have placed bids 
for keywords matching the query and arranges the winners in slots. 
The advertiser pays only if a user clicks on her ad.
Advertiser's bid affects the position of the ad, which in turn affects 
the number of clicks received  and the cost incurred. In addition to the bids, 
the advertiser specifies a daily budget. When the cost charged for the clicks reaches the budget, 
the advertiser's ads stop participating in the auctions.

In what follows, we first model and abstract the budget optimization problem, and then present our stochastic versions, before describing our results.

\subsection{Advertiser's Budget Optimization problem}
We adopt the viewpoint of an advertiser and study the optimization problem she faces. 
The advertiser has to determine the daily budget, a good set of keywords, and 
bids for these keywords so as to  
maximize the effectiveness of her campaign. The daily budget and the choice of 
keywords are business-specific, so they are 
assumed to be given in 
our problem formulation. Effectiveness of a campaign is difficult to quantify  
since clicks  
resulting from some keywords may be more desirable than others, and in some cases, 
just appearing on the results page for a user's query may have some utility. 
For most of the paper, we adopt a common measure of the effectiveness of a
campaign, namely, the {\em number} of clicks obtained\footnote{We 
extend our results to a more general model where clicks for 
different keywords may have different values in Section \ref{sec:clickval}.}.
Further, seen from an individual advertiser's point of view, 
the budgets and bids of other advertisers are fixed for the day. 
We develop most of our discussions assuming that each keyword has a single winning 
bid amount, which is fixed throughout the day and known in advance. This 
models the case of an auction with a single slot, and disregards the 
possibility of other advertisers changing their bids or running out of budget. 
We mention the extensions that eliminate some  
of these assumptions in Section \ref{sec:multslot}. 

Under our assumptions, each keyword $i$ has a single threshold bid amount, 
such that any bid below this amount loses the auction and does not get any 
clicks. Any bid above the threshold wins the auction, and gets clicks with 
cost per click equal to the threshold bid amount\footnote{This assumes {\em 
second-price} auctions, where 
the winner's cost is the highest bid of others. All our results also apply to 
weighted second-price auctions, which are common in search-based advertising.}. 
In this case the advertiser's decision for each keyword becomes binary: 
whether or not to bid on it above its threshold. 
We use decision variables $b_i$, which can be integral ($b_i\in \{0,1\}$) or 
fractional ($b_i \in [0,1]$), to indicate whether or not there is a bid on keyword $i$. 
A fractional bid represents bidding for $b_i$ fraction of the queries that 
correspond to keyword $i$, or equivalently bids on each such query with 
probability $b_i$.  
Integer bid solutions are slightly simpler to implement than  fractional bids 
and are more desirable when they exist. 

Finally, consider the effect of user behavior on the advertiser. 
We abstract it using the function $\cl_i$,
which is the number of clicks the advertiser gets for queries corresponding to keyword $i$.
Each such click entails a cost $\cpc_i$, which is assumed to be known. 
Now, the advertiser is {\em budget-constrained}, 
and some solutions may run out of budget, which decreases the total number of 
clicks obtained. In particular, the advertiser has a global daily budget $B$, which 
is used to get clicks for all of the keywords. When the budget is spent, the 
ads stop being shown, and no more clicks can be bought. We model the limited 
budget as follows.
Consider a solution $\sol$ that bids on some keywords. If the budget  were 
unlimited, then bidding on those keywords would bring $\cl(\sol)$ clicks, 
which together would cost $\cost(\sol)$. But when the budget $B$ is 
smaller than $\cost(\sol)$, this solution runs out of money before the end of 
the day, and misses the clicks that come after that point.  If we assume that 
the queries and clicks for all keywords are distributed uniformly throughout 
the day and are well-mixed, then this solution reaches the budget after  
$B/\cost(\sol)$ fraction of the day passes, missing $(1-B/\cost(\sol))$ 
fraction of the possible clicks for each keyword. As a result, the number of 
clicks collected before the budget is exceeded is 
$\frac{\cl(\sol)}{\cost(\sol)/B}$ in expectation. 

Based on the discussion so 
far, we can now state the optimization problem an advertiser faces.  

\begin{definition} \label{defn:bo}
{\sc Budget Optimization Problem} (BO).
An advertiser has a set $T$ of keywords, with $|T|=n$, and a budget $B$. For each keyword $i\in T$, 
we are given $\cl_i$, the number of clicks that correspond to $i$,
and $\cpc_i$, the cost per click of these clicks. We define $\cost_i = \cpc_i \cdot \cl_i$.
The objective is to find a solution $\sol = (b_1, ..., b_n)$ with a bid 
$0\leq b_i\leq 1$ for each $i\in T$ to maximize 
\begin{equation}\label{objective}
\val(\sol) = \frac{\sum_{i\in T} b_i \cl_i}{\max\left(1, \sum_{i\in T} b_i \cost_i/B\right)}.
\end{equation}
\end{definition}

The numerator of the objective function is the number of clicks available to $\sol$, and the denominator scales it down in the case that the budget is exceeded. If we define $\cl(\sol) = \sum_{i\in T} b_i \cl_i$, $\cost(\sol) = \sum_{i\in T} b_i \cost_i$, and the average cost per click of solution $\sol$ as $\cpc(\sol) = \frac{\cost(\sol)}{\cl(\sol)}$, then 
\begin{equation}\label{altobj}
\val(\sol) = \begin{cases}
\cl(\sol) & \text{if} ~~ \cost(\sol)\leq B\\
B/\cpc(\sol) & \text{if} ~~ \cost(\sol) > B
\end{cases}
\end{equation}
So maximizing $\val(\sol)$ is equivalent to maximizing the number of clicks in case that we are under budget, and minimizing the average cost per click if we are over budget.
We always assume that the keywords are numbered in the order of non-decreasing 
$\cpc_i$, i.e. $\cpc_1\leq \cpc_2 \leq \dots \leq \cpc_n$.

\subsection{Stochastic versions}
Many variables affect the number of clicks that an advertiser receives in a day. 
Besides the advertiser's choice of her own budget and 
keywords which we take to be given, and the choices of other advertisers which remain fixed, 
the main {\em variable} in our problem is the number of queries of relevance 
that users issue on that day, and the frequency with which the ads  
are clicked.\footnote{The nature and number of queries vary significantly. An example in Google Trends  shows the 
spikes in searches for shoes, flowers and chocolate: 
\url{http://www.google.com/trends?q=shoes,flowers,chocolate}.} 
These quantities are not known precisely in advance.
Our premise is that Internet search companies can 
analyze past data and provide probability distributions for parameters of interest. 
They currently do provide limited amount of information about the range of
values taken by these parameters.\footnote{See for example
the information provided to any AdWords advertiser. See also
\url{https://adwords.google.com/support}.}
This motivates us to study the problem in the \emph{stochastic} setting where
the goal is to maximize the {\em expected} value of the objective under such probability
distributions. 

In the stochastic versions of our problem, the set of keywords $T$, the budget 
$B$, and the cost per click $\cpc_i$ for each keyword are fixed and given, 
just like in the BO problem of Definition \ref{defn:bo}. What is different is 
that the numbers of clicks $\cl_i$ corresponding to different keywords are 
random variables having some joint probability distribution. But because general joint probability
distributions are difficult to represent and to work with, we formulate the 
following natural stochastic models. 
(In contrast, the problem where $\cl_i$ are known precisely
for all $i$ is called the {\em fixed} model from here on.)

\begin{description}
\item[Proportional Model]
~The relative proportions of clicks for different keywords remain 
constant. This is modeled by one global random variable for the total 
number of clicks in the day, and a fixed known multiplier for each 
keyword that represents that keyword's share of the clicks.

\item[Independent Keywords Model]
~Each keyword comes with its own probability distribution for the number of clicks, and the samples are drawn from these distributions independently. 

\item[Scenario Model]
~There is an explicit list of $N$
\emph{scenarios}. Each scenario specifies the  
number of clicks 
for each keyword, and has a probability of occurring. We think of 
$N$ as reasonably small, and allow the running time
of algorithms to depend (polynomially) on $N$.
\end{description}

The scenario model is important for two reasons. For one, market
analysts often think of uncertainty by explicitly creating a set of a
few model scenarios, possibly attaching a weight to each scenario. The
second reason is that the scenario model gives us an important segue
into understanding the fully general problem with arbitrary joint
distributions.
Allowing the full generality of an arbitrary joint distribution gives
us significant modeling power, but poses challenges to the algorithm
designer. Since a naive explicit representation of the joint
distribution requires space exponential in the number of random
variables, one often represents the distribution implicitly by a
sampling oracle. A common technique, Sampled Average Approximation
(SAA), is to replace the true distribution $\cal{D}$ by a uniform or
non-uniform distribution $\hat{\cal{D}}$ over a set of samples drawn
by some process from the sampling oracle, effectively reducing the
problem to the scenario model. For some classes of
problems, see e.g. \cite{kleywegt:shapiro,charikar:chekuri:pal,swamy:shmoys:2stage}, it is known that SAA approximates the original
distribution to within an arbitrarily small error using polynomially
many samples. While we are not aware of such bounds applicable to the
budget optimization problem, understanding the scenario model is still
an important step in understanding the general problem.

There are two issues that arise in each of the three stochastic models. 
\begin{itemize}
\item
{\sc Stochastic Evaluation Problem (SE).} 
Given a solution $\sol$, can we evaluate $E[\val(\sol)]$ for the three models
above? 
Even this is nontrivial as is typical in stochastic optimization
problems. 
 It is also of interest in solving the budget optimization problem below.

\item
{\sc Stochastic Budget Optimization Problem} (SBO). 
This is the Budget Optimization problem with one of the stochastic models above determining $\cl_i$ for each $i$, with the objective to maximize 
\begin{equation}\label{expobjective}
E[\val(\sol)] = E \left[ \frac{\sum_{i\in T} b_i \cl_i}{\max\left(1, \sum_{i\in T} b_i \cost_i /B\right)}\right].
\end{equation}
The expectation is taken over the joint distribution of $\cl_i$ for all $i\in T$. 

\end{itemize}

\subsection{Our results}
We present algorithmic and complexity-theoretic results for the  
SE and SBO problems.

For SE problems, our results are as follows. 
The problem is straight-forward to solve for the fixed and scenario models since the expression for the expected value of the objective can be explicitly written in polynomial time. 
For the proportional model, we give an exact algorithm to evaluate a solution, 
assuming that some elementary quantities (such as probability of a range of 
values) can be extracted from the given probability distribution in polynomial 
time. For the independent model, the number of 
possibilities for different click quantities may be exponential in the number 
of keywords, and the problem of evaluating a solution is 
likely to be $\#P$-hard. We give a PTAS for this case. 
These evaluation results are used to derive algorithms for the SBO 
problem, though they may be of independent interest.

Our main results are for the SBO problem. In fact, all our algorithms produce a special kind of solutions called 
\emph{prefix solutions}. A prefix solution bids on some prefix of the list of 
keywords sorted in the increasing order of cost per click ($\cpc_i$), i.e., on the cheap ones. Formally, an \emph{integer prefix 
solution} with bids $b_i$ has the property that there exists some $i^*$ such 
that $b_i = 1$ for all $i\leq i^*$, and $b_i = 0$ for $i > i^*$. For a  
\emph{fractional prefix solution}, there exists an $i^*$ such that $b_i = 1$ 
for $i < i^*$,  $b_i = 0$ for $i > i^*$, and $b_{i^*} \in [0,1]$.  We show:
\begin{itemize}
\item
For the proportional model, we can find an optimal fractional solution in polynomial time if the 
distribution of clicks can be described using polynomial number of points; else, we obtain a PTAS. 
We get this result by showing that the optimal fractional solution in this case
is a prefix solution and giving an algorithm to find the best prefix. 

\item
  Our main technical contribution is the result for the independent model, 
where we prove that every integer solution can be transformed to a prefix 
solution by  
removing a set of expensive keywords and adding a set of cheap ones,  while 
losing at most half of the value of the solution. Thus, some integer prefix is always 
a 2-approximate integer solution. When combined with our PTAS for the evaluation problem, this leads 
  to a $2+\varepsilon$ approximation algorithm. We also show that the best 
  fractional prefix is not in general the optimal fractional solution in this case.
\item
For the scenario model, we show a negative result that finding the optimum, fractional
or integer, is NP-hard. 
In this case, the best prefix solution is arbitrarily far from the optimum. 
\end{itemize}

\subsection{Related work}
Together, our results represent a new theoretical study of stochastic
versions of budget optimization problems in search-related advertising. 
The budget optimization problem was
studied recently~\cite{feldman:budgetoptimization} in the fixed model, when $\cl_i$'s are known. On one hand, our study is more
general, with the emphasis on the uncertainty in modeling $\cl_i$'s and the stochastic models we have formulated. We do not know of prior
work in this area that formulates and uses our stochastic models.
On the other hand, our study is less general as it does not consider the 
interaction between keywords that occurs when a user's search query matches two 
or more keywords, which is studied in~\cite{feldman:budgetoptimization}. 

Stochastic versions of many optimization problems have been considered,
 such as facility location, Steiner trees, bin-packing and LP (see, for 
example, the survey~\cite{swamy:shmoys:2stage}).  
Perhaps the most relevant to our setting is the work on the stochastic knapsack 
problem, of which several versions have been studied. 
Dean~et~al.~\cite{dean:goemans:vondrak}   
consider a version of the problem in which item values are 
fixed, and item sizes are independent random variables.  The realization of 
an item's size becomes known once it is placed in the knapsack, so an 
algorithm has to select items one at a time, until the knapsack capacity is 
exceeded.   In \cite{kleinberg:rabani:tardos} and 
\cite{goel:indyk}, a version of the problem with fixed item values and random sizes 
is considered as well, but there the goal is to choose a valuable set of items 
whose probability of exceeding the knapsack capacity is small.
Other authors \cite{carraway:schmidt,henig,sniedovich,steinberg:parks}  have 
studied versions with fixed item sizes but random values. If viewed as a 
version of stochastic knapsack, our problem is different 
from all of these in several respects. First, there is no hard capacity 
constraint, but instead the objective function decreases continuously if the 
cost of the keywords (which is analogous to the size of the items) exceeds the 
budget (the analog of the knapsack capacity). The second difference is that 
in our model, both the number of clicks and the cost of the keywords (i.e. 
item values and sizes, respectively) are random, but their ratio for each 
particular keyword (item) is fixed and known. Another difference is that 
previous work on stochastic knapsack considers independent distributions of 
item parameters, whereas two of our models (proportional and scenario) have 
correlated variables. Furthermore, although the greedy 
algorithm which takes items in the order of their value-to-size ratio is 
well-known and variations of it have been applied to knapsack-like 
problems, our analysis proving the 2-approximation result is new.

Recently, Chakrabarty et al.\ \cite{chakrabarty:zhou:lukose} considered an 
online knapsack problem with the assumption of small element sizes, and 
Babaioff et al.\ \cite{babaioff:knapsack} considered an online  
knapsack problem with a random order of element arrival, both  
motivated by bidding in advertising auctions.  
The difference with our work is that these authors consider the problem in the online 
algorithms framework, and analyze the competitive ratios of the obtained 
algorithms. In contrast, our algorithms make decisions offline, and we analyze 
the obtained approximation ratios for the expected  
value of the objective. Also, our algorithms base their decisions on the 
probability distributions of the clicks, whereas the authors of 
\cite{babaioff:knapsack} and \cite{chakrabarty:zhou:lukose} do not assume any 
advance knowledge of these  
distributions. The two approaches are in some sense complementary: online 
algorithms have the disadvantage that in practice it may not be possible to 
make new decisions about bidding every time that a query arrives, and 
stochastic optimization has the disadvantage of requiring the 
knowledge of the probability distributions.

Also motivated by advertising in search-based auctions, Rusmevichientong and 
Williamson \cite{rusmevichientong:williamson} have studied the {\em keyword 
selection} problem, where the goal is to select a subset of keywords from a 
large pool for the advertiser to choose to bid.  
Their model is similar to our proportional 
model,
but the proportions of clicks for different keywords
are unknown. An adaptive algorithm is developed that learns the proportions by 
bidding on different prefix solutions, and eventually converges to 
near-optimal profits~\cite{rusmevichientong:williamson}, 
assuming that various parameters are concentrated around their means. The 
difference with our work is that we consider algorithms that solve the problem 
in advance, and not by adaptive learning, and work for any arbitrary (but 
pre-specified) probability  distributions. 

There has been a lot of other work on search-related auctions 
 in the presence of budgets, but it has primarily
focused on the game-theoretic aspects~\cite{ostrovsky:edelman:schwartz,aggarwal:goel:motwani}, strategy-proof mechanisms~\cite{borgs:multiunit,borgs:bidoptimization}, and revenue
maximization~\cite{mehta:saberi,mahdian:nazerzadeh:saberi}.

\subsection{Map}
We briefly discuss the fixed case first, and then focus on the three stochastic models
in the following sections; in each case, we solve both evaluation and BO problems. 
Finally, we present some extensions of our work and 
state a few open problems. Some of the proofs appear in the Appendix.


\section{Fixed  model}\label{fixedsec}
For the BO problem in the fixed model, a certain fractional prefix, which is easy to find, 
is the optimal solution.  
The algorithm is analogous to that for the fractional knapsack problem. We find the maximum index $i^*$ such that $\sum_{i\leq i^*} \cost_i \leq B$. If $i^*$ is the last index in $T$, we set $b_i = 1$ for all keywords $i$. Otherwise find a fraction $\alpha \in [0,1)$ such that $\sum_{i\leq i^*} \cost_i + \alpha\cdot \cost_{i^*+1} = B$, and set $b_i = 1$ for $i\leq i^*$, $b_{i^*+1} = \alpha$, and $b_i = 0$ for $i>i^*+1$. 

\begin{theorem}\label{fixedthm}
In the fixed model, the optimal fractional solution for the BO problem 
is the maximal prefix whose cost does not exceed the budget, which can be found in linear time.
\end{theorem}

\noindent The integer version of this problem is NP-hard by reduction from \textsc{knapsack}.


\section{Proportional model}\label{interchangearg}

In the proportional model of SBO, we are given $q_i$, the \emph{click frequency} for each keyword $i\in T$, with $\sum_{i\in T} q_i = 1$. The total number of clicks is denoted by a random variable $C$, and has a known probability distribution $p$.  The number of clicks for a keyword $i$ is then determined as $\cl_i = q_i \cdot C$. For a specific  value $c$ of $C$, let $\cl_i^c = q_i c$ and $\cost_i^c = \cpc_i \cl_i^c$.
The objective is to maximize the expected number of clicks, given by expression (\ref{expobjective}).

\begin{theorem}\label{prefixoptprop}
The optimal fractional solution for the SBO problem in the proportional model 
is a fractional prefix solution.
\end{theorem}

\noindent The proof is by an interchange argument and appears in Section 
\ref{sec:intarg}.  
We now show how to solve the evaluation problem efficiently in the 
proportional model, and then use it to find the best prefix, which by Theorem 
\ref{prefixoptprop} is the optimal fractional solution 
to SBO.

\subsection{Evaluating a solution}
Assuming that the distribution for $C$ is given in such a way that it is easy 
to evaluate $\Pr[C>c^*]$ and $\sum_{c\leq c^*}c\cdot p(c)$ for any $c^*$, 
we show how to find $E[\val(\sol)]$ for any given solution $\sol$ without 
explicitly going through all possible values of $C$ and evaluating the 
objective function for each one. 

The solution $\sol$ may be under or over budget depending on the value of $C$. Define a threshold  $c^* = B/\sum_{i\in T} b_i q_i \cpc_i$, so that for $c\leq c^*$, $\cost^c(\sol) \leq B$, and for $c>c^*$, $\cost^c(\sol) > B$. Notice that in the proportional model, $\cpc(\sol)$ is independent of $C$, as both $\cl(\sol)$ and $\cost(\sol)$ are proportional to $C$. Then using expression (\ref{altobj}) for $\val(\sol)$, the objective becomes easy to evaluate:
\begin{equation}\label{propobj}
E[\val(\sol)] = \sum_{i\in T} b_i q_i \sum_{c\leq c^*}c~p(c) + \frac{B}{\cpc(\sol)} \Pr[C>c^*].
\end{equation}

\subsection{Finding the optimal prefix}
It is nontrivial to find the best fractional prefix solution for the proportional
case, and we mention two approaches that do not work. 
%
One simple way to find a prefix in the proportional model is to convert it to a fixed
case problem by setting the number of clicks for a keyword to its expectation. 
This approach fails on an example with two keywords and two possible values of 
$C$: a likely value of zero and a small-probability large value.
Another approach is some greedy procedure 
that lengthens the prefix while the solution improves.  This does not work 
either, because the expected value of the solution as a function of the length 
of the prefix can have multiple local maxima. 

The best prefix can be found by producing a list of  $O(n+t)$ prefixes (out of 
uncountably many possible ones) 
containing the optimum, as explained in Section \ref{sec:optprefprop}. Here $t$ 
is the number of possible  
values of $C$.  If $t$ is not polynomial in $n$, the probability that $C$ 
falls between successive powers of  
$(1+\varepsilon)$ can be combined into buckets, yielding a PTAS for 
the problem.

\begin{theorem}\label{thm:findoptprop}
The optimal fractional solution to SBO problem in the proportional model can be found exactly in time $O(n+t)$, where $t$ is the number of possible values of $C$, or approximated by a PTAS.
\end{theorem}


\section{Independent model}
In the independent model of SBO, the number of clicks for keyword $i\in T$, 
$\cl_i$,  has a probability distribution $p_i$ (which can be different for 
different keywords). The key distinguishing feature of this model is that for 
$i\neq j$, the variables $\cl_i$ and $\cl_j$ are independent. 
This model is more complex than the ones discussed so far. A three-keyword 
example in Section \ref{sec:optnotpref} shows the following.

\begin{theorem}\label{highvar}
In the independent model of the SBO problem, the optimal fractional solution 
may not be a prefix solution. 
\end{theorem}

However, in Section 
\ref{2approx} we prove that some integer prefix solution is a 2-approximate 
integer  solution. But finding the best integer prefix requires the ability to evaluate a  
given solution, which in this model is likely to be $\#P$-hard. We develop 
a PTAS, based on dynamic programming, for evaluating a solution 
(the algorithm is presented in Section \ref{indepevalptas}). 
Combined, these two results imply a $(2+\varepsilon)$-approximation 
for the SBO problem in the independent model.


\subsection{Prefix is a 2-approximation}\label{2approx}
In this section we show that for any instance of the SBO problem in the 
independent model, there exists an integer prefix solution whose expected 
value is at least half that of the optimal integer solution. 
In particular, any integer solution $\sol$ can be transformed 
into a prefix solution $\solv$ without losing more than half of its value. 
Let $S = \{i ~|~b_i = 1\}$ be the set of keywords that $\sol$ bids on. 

We make some definitions that allow us to specify the prefix solution $\solv$ 
precisely in Theorem \ref{2approxthm}. Let $\sigma$ be the event that clicks for each keyword $i\in T$ 
come in quantity $\cl^{\sigma}(i)$.  Then its probability is 
$p(\sigma) = \prod_{i\in T} p_i(\cl^{\sigma}(i)).$
Define 
the number of clicks available to solution $\sol$ in the event $\sigma$ as 
$\cl^{\sigma}(\sol) = \sum_{i\in S} \cl^{\sigma}(i)$, the corresponding cost 
per keyword 
$\cost^{\sigma}(i) = \cpc_i \cdot \cl^{\sigma}(i)$, and total cost
$\cost^{\sigma}(\sol) = \sum_{i\in S} \cost^{\sigma}(i)$.
The effective number of clicks (after taking the budget into account) that solution $\sol$ gets from keyword $i$ in the event $\sigma$ is
$$\overline{\cl}_{S}^{\sigma}(i) = \frac{\cl^{\sigma}(i)}{\max (1, \cost^{\sigma}(\sol)/B)},$$
and the total effective number of clicks is 
$\overline{\cl}^{\sigma}(\sol) = \sum_{i \in S} \overline{\cl}_S^{\sigma}(i)$. 
Then the objective becomes the sum of effective number of clicks in each 
scenario, weighted by that scenario's probability: 
$E[\val(\sol)] = \sum_{\sigma} p(\sigma) \overline{\cl}^{\sigma}(\sol)$.

\smallskip
Let $i^*(\sol)$ be the minimum index $i^*$ such that keywords up to $i^*$ 
account for half the clicks:
$$\sum_{\sigma} p(\sigma) \sum_{i\in S, i\leq i^*} \overline{\cl}_S^{\sigma}(i) \geq \frac{1}{2}~E[\val(\sol)].$$

\begin{theorem}\label{2approxthm}
For any integer solution $\sol$ to the SBO problem with independent keywords, there exists an integer prefix solution $\solv$ such that $E[\val(\solv)] \geq \frac{1}{2}~E[\val(\sol)]$.  In particular, the solution $\solv$ bidding on the set $V = \{i~|~i\leq i^*(\sol)\}$ has this property.
\end{theorem}

The idea of the proof is to think of the above prefix solution as being 
obtained in two steps from the original solution $\sol$. First, we truncate 
$\sol$ by discarding all keywords after $i^*$. Then we fill in the gaps in the 
resulting solution in order to make it into a prefix. To analyze the result, 
we first show that all keywords up to $i^*$ are relatively cheap, and that the 
truncated solution (called $\solu$) retains at least half the value of the 
original one (Claim \ref{valU}). Then we show that filling in the gaps 
preserves this guarantee. Intuitively, two good things may happen: either 
clicks for the new keywords don't come, in which case we get all the clicks we 
had before; or they come in large quantity, spending the budget, which is good 
because they are cheap. Lemma \ref{onescen} analyzes what happens if new 
clicks spend $\alpha^\sigma$ fraction of the budget.

Let $i^* = i^*(\sol)$. To analyze our proposed prefix solution $\solv$, we break the set $V$ into two disjoint sets $U$ and $N$.  $U = V \cap S = \{i\leq i^* ~|~i\in S \}$ is the set of cheapest keywords that get half the clicks of $\sol$. The new set $N = V\setminus S = \{i~\leq~i^*~|~i\notin S\}$ fills in the gaps in $U$. Let $\solu$ and $\soln$ be the solutions that bid on keywords in $U$ and $N$ respectively.

Define the average cost per click of solution $\sol$ as
$$\cpc^* = \frac{\sum_{\sigma} p(\sigma)~ \sum_{i\in S} \cpc_i~ \overline{\cl}_{S}^{\sigma}(i)}{\sum_{\sigma} p(\sigma) ~\overline{\cl}^{\sigma}(\sol)},$$
where the numerator is the average amount of money spent by $\sol$, and the 
denominator is the average number of clicks obtained.  A useful fact to notice 
is that since the numerator of this expression never exceeds the budget, and 
the denominator is equal to $E[\val(\sol)]$, we have that  
\begin{equation}\label{valcpc}
E[\val(\sol)] \leq \frac{B}{\cpc^*}.
\end{equation}

We make two observations about $\solu$ and $i^*$.

\begin{claim}\label{valU}\label{cpcclaim}
$E[\val(\solu)] ~\geq~ \frac{1}{2} E[\val(\sol)]$ ~and~ $\cpc_{i^*} ~\leq~ 2\cpc^*$.
\end{claim}
\begin{proof}
A keyword in $U$ does not necessarily contribute exactly the same number of clicks when it appears as part of solution $U$ as it does when it is part of the solution $S$. However, in $U$ it contributes at least as much.  Formally, observe that 
since $U\subseteq S$, for all $\sigma$, $cost^\sigma (\solu) \leq cost^\sigma (\sol)$, which implies that $\overline{clicks}_{U}^{\sigma}(i) \geq \overline{clicks}_{S}^{\sigma}(i)$ for any $i\in U$.  
So $$E[\val(\solu)] = 
\sum_{\sigma} p(\sigma) \sum_{i \in U} \overline{clicks}_{U}^{\sigma}(i) \geq \sum_{\sigma} p(\sigma) \sum_{i \in U} \overline{clicks}_{S}^{\sigma}(i)
 \geq \frac{1}{2}~E[\val(\sol)]$$ by definitions of $U$ and $i^*$.

The second part follows by Markov's inequality:
$$\cpc^* \geq \frac{\sum_{\sigma} p(\sigma)~ \sum_{i\in S, i\geq i^*} \cpc_{i^*}~ \overline{\cl}_{S}^{\sigma}(i)}{\sum_{\sigma} p(\sigma) ~\overline{\cl}^{\sigma}(\sol)} \geq \cpc_{i^*} \cdot \frac{1}{2},$$ where the second inequality is by minimality of $i^*$.
\end{proof}

Now we state the main lemma.

\begin{lemma}\label{onescen}
For any $\sigma$, let $\alpha^\sigma  = \min(B, ~\cost^\sigma(\soln))/B$. Then
$$\overline{\cl}^\sigma (\solv) ~\geq~ \alpha^\sigma  \frac{B}{2 \cpc^*} + (1-\alpha^\sigma )~ \overline{\cl}^\sigma (\solu).$$
\end{lemma}
\begin{proof}
The idea here is that $\alpha^\sigma$ is the fraction of the budget spent by 
the new keywords (ones from set $N$) in the event $\sigma$. So 
$(1-\alpha^\sigma)$ fraction of the budget can be used to buy 
$(1-\alpha^\sigma)$ fraction of clicks that $\solu$ was getting, and 
$\alpha^\sigma$ fraction is spent on keywords (whether from $U$ or $N$) that 
cost at most $2 \cpc^*$. A more formal analysis follows.

If $\cost^\sigma(\soln) \geq B$, then $\cost^\sigma(\solv) \geq B$, so
\begin{eqnarray*}
\overline{\cl}^\sigma (\solv) =
\frac{\sum_{i\in V} \cl^\sigma (i)}{\cost^\sigma(\solv)/B}  
 = B\cdot \frac{\sum_{i\in V} \cl^\sigma (i)}{\sum_{i\in V} \cpc_i \cl^\sigma (i)} \geq \frac{B}{\cpc_{i^*}}\geq \frac{B}{2 \cpc^*},
\end{eqnarray*}
which proves the lemma for the case of $\alpha^\sigma =1$. Intuitively, in this case the whole budget is spent, and since all keywords in $V$ cost at most $2 \cpc^*$ (by Claim \ref{valU}), $V$ gets at least $\frac{B}{2\cpc^*}$ clicks.
For the rest of the proof assume that $\cost^\sigma(\soln) < B$. Then $\alpha^\sigma  = \frac{\cost^\sigma(\soln)}{B}<1$. 

Another simple case is  $\cost^\sigma(\solv) \leq B$.  Then the budget is not reached and $V$ collects all the clicks from $U$ and $N$: 
$$
\overline{\cl}^\sigma (\solv) = \cl^\sigma(\soln) + \cl^\sigma(\solu) \geq 
\frac{\cost^\sigma(\soln)}{2\cpc^*} + \overline{\cl}^\sigma(\solu) \geq \frac{\alpha^\sigma B}{2 \cpc^*} + (1-\alpha^\sigma )~ \overline{\cl}^\sigma (\solu).
$$
Now consider the case when $\cost^\sigma(\soln) + \cost^\sigma(\solu) > B$.  
Here at most $\alpha^\sigma$ fraction of the budget is used for the new keywords from $N$, which cost at most $2 \cpc^*$ per click, and the remaining budget is able to buy $(1-\alpha^\sigma)$ fraction of the clicks that $\solu$ was getting.

Define $\cpc^\sigma(\solu) = \frac{\cost^\sigma(\solu)}{\cl^\sigma(\solu)}$, and similarly for $N$. Then 
\begin{equation}\label{cpcU}
\frac{B}{\cpc^\sigma(\solu)} ~=~ \frac{\cl^\sigma(\solu)}{\cost^\sigma(\solu)/B} ~\geq~ \overline{\cl}^\sigma(\solu).
\end{equation}
\begin{eqnarray*}
\textrm{Now,~}~\overline{\cl}^\sigma (\solv) 
&=& \frac{\cost^\sigma(\solu)}{\cost^\sigma(\solv)}\cdot\frac{B}{\cpc^\sigma(\solu)} + \frac{\cost^\sigma(\soln)}{\cost^\sigma(\solv)}\cdot\frac{B}{\cpc^\sigma(\soln)} \\ 
&=& \frac{(1-\alpha)B}{\cpc^\sigma(\solu)} + \left[\frac{\cost^\sigma(\solu)}{\cost^\sigma(\solv)} - (1-\alpha)\right] \frac{B}{\cpc^\sigma(\solu)} + ~\frac{\cost^\sigma(\soln)}{\cost^\sigma(\solv)}\cdot\frac{B}{\cpc^\sigma(\soln)} \\ 
&\geq& \frac{(1-\alpha)B}{\cpc^\sigma(\solu)} + \left[\frac{\cost^\sigma(\solu)}{\cost^\sigma(\solv)} - (1-\alpha)\right] \frac{B}{2\cpc^*} 
 + ~\frac{\cost^\sigma(\soln)}{\cost^\sigma(\solv)}\cdot\frac{B}{2\cpc^*} \\
&\geq& (1-\alpha)\overline{\cl}^\sigma(\solu) + \alpha \frac{B}{2\cpc^*},
\end{eqnarray*}
where the first inequality follows because $\cpc^\sigma(\solu) \leq 2\cpc^*$, $\cpc^\sigma(\soln) \leq 2\cpc^*$, and the quantity in square brackets is non-negative. The second inequality follows from (\ref{cpcU}).
\end{proof}

\begin{proofof}{Theorem \ref{2approxthm}}
We now use the above results to prove the theorem.  
Let $\sigma_U$ be the event that clicks for each keyword $i\in U$ come in 
quantity $\cl^{\sigma_U}(i)$, whose probability is 
$p(\sigma_U) = \prod_{i\in U} p_i(\cl^{\sigma_U}(i))$.
Here the independence of keywords becomes crucial.  In particular, what we need is that the number of clicks that come for keywords in $U$ is independent of the number of clicks for keywords in $N$. So the probability of $\sigma_V$ is the product of $p(\sigma_U)$ and $p(\sigma_N)$, where $\sigma_V$ is the event that both $\sigma_U$ and $\sigma_N$ happen.
Notice that $\alpha^\sigma$ of Lemma \ref{onescen} depends only on keywords in $N$, and  is independent of what happens with keywords in $U$. So here we call it $\alpha^{\sigma_N}$. 
We have

\begin{eqnarray*}
E[\val(\solv)] &=& \sum_{\sigma_V} p(\sigma_V) \overline{\cl}^{\sigma_V} (\solv) \\
&\geq& \sum_{\sigma_N} \sum_{\sigma_U} p(\sigma_N)p(\sigma_U) 
\left[\frac{\alpha^{\sigma_N} B}{2 \cpc^*} + (1-\alpha^{\sigma_N} )~  
\overline{\cl}^{\sigma_V}(\solu)\right] \\
&=&
 \sum_{\sigma_N} p(\sigma_N) \left[\frac{\alpha^{\sigma_N} B}{2 \cpc^*} +  (1-\alpha^{\sigma_N} ) \sum_{\sigma_U} p(\sigma_U) 
 \overline{\cl}^{\sigma_V}(\solu)\right]  \\
&\geq& \sum_{\sigma_N} p(\sigma_N) ~\frac{1}{2}E[\val(\sol)]  ~=~ \frac{1}{2}E[\val(\sol)],
\end{eqnarray*}

\noindent bounding both $\frac{B}{2\cpc^*}$ and $E[\val(\solu)]$ by 
$\frac{1}{2}E[\val(\sol)]$ using inequality (\ref{valcpc}) and Claim~\ref{valU}. 
\end{proofof} 

Theorem \ref{2approxthm} combined with PTAS for the evaluation problem in the 
independent model (which 
is presented in Section \ref{indepevalptas} of the Appendix) gives a simple 
algorithm for finding a $(2+\varepsilon)$-approximate  
solution: evaluate each integer prefix using the PTAS and output the one with 
maximum value.

\begin{theorem}
There is a $(2+\varepsilon)$-approximation algorithm for the SBO problem in 
the independent model, which runs in time polynomial in $n$, 
$\frac{1}{\varepsilon}$, and $\log M$, where $M$ is the maximum possible cost 
of all clicks.
\end{theorem}


\section{Scenario model}

In the scenario model, we are given $T$, $B$ and costs $\cpc_i$ as usual. The 
numbers of clicks are determined by a set of scenarios $\Sigma$ and a 
probability distribution $p$ over it, so that a scenario $\sigma \in \Sigma$ 
materializes with probability $p(\sigma)$, in which case each keyword $i$ gets 
$\cl^\sigma_i$ clicks. The scenarios are disjoint and 
$\sum_{\sigma\in \Sigma} p(\sigma) = 1$. The reason this model does not 
capture the full generality of arbitrary distributions is that we assume that 
the number of scenarios, $|\Sigma|$, is relatively small, in the sense that 
algorithms are allowed to run in time polynomial in $|\Sigma|$. On the other 
hand, if, for example, we express the independent model in terms of scenarios, 
their number would be exponential in the number of keywords. 

The evaluation of a given solution in the scenario model does not present a 
problem, as it can be done explicitly in time polynomial in $|\Sigma|$, by 
evaluating each scenario and taking the expectation. Nevertheless, this is the 
most difficult model for SBO that we consider.  
We show two negative results. 

\begin{theorem}\label{nphardthm}
The SBO problem is NP-hard in the scenario model.
\end{theorem}

\begin{theorem}\label{badprefix}
The gap between the optimal fractional prefix solution and the optimal (integer or fractional) solution to the SBO problem in the scenario model can be arbitrarily large.
\end{theorem}

The proof of Theorem \ref{nphardthm} shows, by reduction from {\sc clique}, 
that it is NP-hard to find either an integer or a fractional solution to this 
problem. Proofs of the two theorems appear in Sections \ref{nphard} and 
\ref{sec:gappf}, respectively.



\section{Extension to click values}\label{sec:clickval}
Here we show that our results easily generalize to the case when the clicks from different keywords have different values to the advertiser.  For example, a weight associated with a keyword might represent the probability that a user clicking on the ad for that keyword will make a purchase. 

For each keyword $i$, we are given a weight $w_i$ which is the value of a click associated with this keyword, and we would like to maximize the weighted number of clicks obtained:
$$E[\val(\sol)] = E \left[ \frac{\sum_{i\in T} b_i w_i \cl_i}{\max\left(1, \sum_{i\in T} b_i \cpc_i \cl_i/B\right)}\right].$$
Obviously, the keywords with $w_i=0$ can be just discarded. Now we make a substitution of variables, defining $\cl'_i = w_i \cl_i$ and $\cpc'_i = \cpc_i/w_i$. Substituting them into the objective function,
$$E[\val(\sol)] = E \left[ \frac{\sum_{i\in T} b_i \cl'_i}{\max\left(1, \sum_{i\in T} b_i \cpc'_i \cl'_i/B\right)}\right],$$
we see that the problem reduces to the original unweighted SBO instance, with different keyword parameters. The proportional, independent, and scenario models of click arrival maintain their properties under this transformation, only some of the distributions for the numbers of clicks have to be scaled.

\section{Concluding Remarks}\label{concl}
We have initiated the study of stochastic version of budget optimization.
We obtained approximation results via prefix bids and showed hardness results 
for other cases. A lot remains to be done, both technically and conceptually. 
Technically,  
we need to extend the results to the case when there are interactions between 
keywords, that is, two or more of them apply to a user query and some resolution
is needed. Also, we need 
to study online algorithms, including online budget optimization. Further,
we would like to obtain some positive approximation results for the 
scenario model, which seems quite intriguing from an application point of view. 
The conceptual challenge is one of modeling. Are there other suitable 
stochastic models for search-related advertising, that are both expressive,
physically realistic and computationally feasible? 

\section{Acknowledgements}
We thank Jon Feldman, Anastasios Sidiropoulos, and Cliff Stein for helpful discussions.

\pagebreak
{\small
\bibliographystyle{plain}
\bibliography{../../bib/names,../../bib/conferences,../../bib/bibliography}
}

\pagebreak
\appendix
\section{Appendix}

\subsection{Proof of Theorem \ref{prefixoptprop}}\label{sec:intarg}
We use an interchange argument to show that any solution can be transformed into a prefix solution without decreasing its value.  Consider a solution $\sol$. If $\sol$ is not a prefix solution, then there exist keywords $i$ and $j$ with $i<j$, $b_i<1$, and $b_j>0$. Choose the smallest such $i$ and the largest such $j$. If $q_i \cpc_i = 0$, set $b_i = 1$ and continue. Otherwise pick the maximum $\delta_i, \delta_j > 0$ that satisfy
$$\delta_i \leq 1-b_i,~~~~~~
\delta_j \leq b_j,~~~~~~~
\delta_i = \frac{q_j \cpc_j}{q_i \cpc_i} ~\delta_j.$$
If we assign $b_i' = b_i + \delta_i$, $b_j' = b_j - \delta_j$, and $b_k'=b_k$ for $k\notin\{i,j\}$, then we get a solution $\sol'$ such that for any $c$, $\cost^c(\sol') = \cost^c(\sol)$ and $\cl^c(\sol') \geq 
\cl^c(\sol)$:
\begin{eqnarray*}
\cost^c(\sol')-\cost^c(\sol) &=& c \cdot (q_i \cpc_i \delta_i - q_j \cpc_j 
\delta_j) ~=~ 0 \\
\cl^c(\sol')- \cl^c(\sol) &=& c \cdot (q_i \delta_i - q_j \delta_j) 
 ~=~ c \cdot q_j (\frac{\cpc_j}{\cpc_i} - 1) \delta_j ~\geq~ 0.\\
\end{eqnarray*}
Since for any $c$, the value of the solution does not decrease, the expected value over $C$ does not decrease either, $E[\val(\sol')] \geq E[\val(\sol)]$. As a result of the transformation, either $b_i'=1$ or $b_j'=0$, so the process terminates after a finite number of steps, resulting in a prefix solution with expected value at least that of the original one.

\subsection{Proof of Theorem \ref{thm:findoptprop}}\label{sec:optprefprop}
We \emph{mark} some points in the space of possible prefixes.  First, we mark all the integer prefixes.  Then, for each  value $c$ of $C$ that has non-zero probability, we mark the threshold prefix $\sol$ that exactly spends the budget for $C=c$, i.e. such that $\cost^c(\sol) = B$.
This partitions the space of prefixes into intervals.  Notice that for any two prefix solutions $\sol$ and $\sol'$ inside of the same interval $I$, the set of values of $C$ that cause these solutions to exceed the budget is the same, i.e. $\{c~|~\cost^c(\sol) > B\} = \{c~|~\cost^c(\sol') > B\}$. Call this set $C_I^>$.

Now we show how to find the optimal prefix solution inside an interval defined by the marked points. Consider such an interval $I$, and suppose that all prefix solutions inside $I$ bid $b_j = 1$ for $j<i$, $b_j = 0$ for $j>i$, and $b_i \in (b_1, b_2)$ for some $0\leq b_1 < b_2 \leq 1$. Then the objective function for solutions in this interval becomes (analogously to equation (\ref{propobj}))
$$\sum_{c\notin C_I^>} c~p(c) \left( \sum_{j<i} q_j + b_i q_i \right) + 
 B \Pr[C \in C_I^>] \cdot 
\frac{\sum_{j<i} q_j + b_i q_i }{\sum_{j<i} q_j \cpc_j + b_i q_i  \cpc_i},
$$
which we have to maximize over the possible values of the variable $b_i$. This 
can be done by taking the derivative of this expression with respect to $b_i$ 
and setting it to zero, which has at most one solution on the interval 
$(b_1, b_2)$. If this solution exists, we add it to a set of 
\emph{interesting} points. To obtain the overall optimum, we evaluate all 
prefixes defined by the $marked$ and $interesting$ points.

\subsection{Proof of Theorem \ref{highvar}}\label{sec:optnotpref}
We give an example with three keywords in which the optimal solution bids on 
the first and third keywords, and gets more clicks than any (even fractional) 
prefix solution. The idea  is that the second and third keywords cost about 
the same, but the third one is better because it always comes in the same 
quantity, whereas the second one has high variance. Let $\cpc_1 = 0$, 
$\cpc_2 = 1$, $\cpc_3 = 1$; $\cl_1 = 1$ with probability 1,  $\cl_2 = (0~or~1)$ 
with probability $\frac{1}{2}$ each, and $\cl_3 = 1$ with probability 1; $B=1$. 
The optimal solution is $b_1 = b_3 = 1$ and $b_2=0$, which always gets 2 
clicks. The best prefix solution is $b_1 = b_2 = b_3 = 1$, which gets 2 or 1.5 
clicks with probability $\frac{1}{2}$ each, or only 1.75 clicks in  
expectation. The example can be modified so that the third keyword is strictly 
more expensive than the second one.

\subsection{Evaluating a solution in the independent model}\label{indepevalptas}
In this section we present a PTAS for the SE problem in the independent model.  
We are given an instance of the SBO problem, and an (integer or fractional) 
solution $\sol$.  
For a keyword $i\in T$, let $C_i = \{c~|~p_i(c)>0\}$ be the set of values that $\cl_i$ can take. For now we assume that $\sum_i |C_i|$ is polynomial in the size of the input, and later show how to remove this assumption. Let $\cost(\sol_{-i}) = \sum_{j\neq i} b_j \cpc_j \cl_j$ be the cost of clicks for all keywords except $i$. By some algebraic manipulation, one can show the following.

\begin{claim} \label{ptasclaim}
$$E[\val(\sol)] = 
\sum_{i\in T} \sum_{c\in C_i} p_i(c) ~b_i c ~ \sum_{d \geq 0} \frac{1}{f_i(c,d)} \Pr[\cost(\sol_{-i}) = d],
$$
\noindent where $f_i(c,d) = \max(1, \frac{d+c \cdot \cpc_i}{B})$. 
\end{claim}
In this expression, $b_i c$ is the number of clicks from keyword $i$, and the expression in the third sum is the amount by which this number should be scaled because of the budget. The variable $d$ represents the cost of all keywords other than $i$.

As a result, the problem of finding $E[\val(\sol)]$ reduces to evaluating, for any given $i$ and $c$, the expression
\begin{equation}\label{ptaseqn}
s(i, c) = \sum_{d \geq 0} \frac{1}{f_i(c,d)} \Pr[\cost(\sol_{-i}) = d].
\end{equation}

\begin{lemma}\label{lemma:indepeval}
For any given $\varepsilon > 0$, there is a polynomial-time algorithm that finds a value $s'$ such that $s(i, c) \leq s' \leq (1+\varepsilon)~s(i, c)$.
\end{lemma}
\begin{proof}
We build a dynamic programming table that represents an estimate of $\Pr[\cost(\sol_{-i}) = d]$ as a function of $d$. Fix an ordering of elements in $T- \{i\}$ and construct a table $P$ indexed by $j$ and $d$, where $P(j,d)$ is the probability that the total cost of the first $j$ elements is $d$. To make sure the table is of polynomial size, scale the costs so that the minimum non-zero value of $\cost_j$ for any $j$ is 1, and restrict the possible values of $d$ to 0 and $(1+\frac{\varepsilon}{n})^k$ for non-negative integers $k$. 
If we let $M = \sum_{i\in T} \max\{c \cdot \cpc_i ~|~c\in C_i\}$ be the maximum possible cost of all the clicks, then the number of values of $d$ in the table is at most $\log_{1+\varepsilon/n} M = O(\frac{n}{\varepsilon} \log M)$, which is polynomial in the size of the input. 

The table is initialized with  $P(0,0)=1$ and other entries equal to zero. Then for each keyword $j\in T- \{i\}$, each possible number of clicks $c\in C_j$, and each entry $P(j-1, d)$ in the previous row, we update $P(j, \lfloor d+c\cdot cpc_j \rfloor) = P(j, \lfloor d+c\cdot cpc_j \rfloor) + p_j(c) \cdot P(j-1, d)$. Here the operator $\lfloor ~ \rfloor$ represents rounding down to the next available value of $d$. After filling the table, the algorithm outputs the value of expression (\ref{ptaseqn}) as determined by probabilities in the last row of the table.

To bound the error incurred by rounding down the costs, we consider an event $\sigma$ that specifies a number of clicks $c_j \in C_j$ for each $j\in T-\{i\}$, and has probability $p(\sigma) = \prod_j p_j(c_j)$. Expression (\ref{ptaseqn}) can be rewritten as 
\begin{equation}\label{ptaseqn1}
s(i,c) = \sum_\sigma p(\sigma) \cdot \frac{1}{f_i(c, \cost^\sigma(\sol_{-i}))}
\end{equation}

As a result of a series of updates, the probability of $\sigma$ contributes to some entry of the last row of $P$, say to the one with $d^\sigma=(1+\frac{\varepsilon}{n})^k$.  This $d^\sigma$ is an estimate of the value of $\cost^\sigma (\sol_{-i}) = \sum_{j\neq i} b_j \cpc_j c_j$. Since we only rounded down, we have $d^\sigma \leq \cost^\sigma (\sol_{-i})$. Now note that since the intervals between successive powers of $(1+\frac{\varepsilon}{n})$ are increasing, the biggest amount that we could have lost during any one rounding is $(1+\frac{\varepsilon}{n})^{k+1} - (1+\frac{\varepsilon}{n})^k = \frac{\varepsilon}{n} ~(1+\frac{\varepsilon}{n})^k = \frac{\varepsilon}{n} \cdot d^\sigma$. Since we performed the rounding during at most $n$ updates relevant to $\sigma$, we have that the true value $\cost^\sigma (\sol_{-i}) \leq d^\sigma + n \cdot \frac{\varepsilon}{n} \cdot d^\sigma = (1+\varepsilon)d^\sigma$. So we have that the estimated cost $d^\sigma$ for the event $\sigma$ is $\frac{\cost^\sigma (\sol_{-i})}{1+\varepsilon} \leq d^\sigma \leq \cost^\sigma (\sol_{-i})$. 

Now the only thing left to do in order to show that the algorithm evaluates expression (\ref{ptaseqn}) accurately is to take into account $f_i(c,d)$. By monotonicity of $f_i$, we have $$f_i\left(c,\frac{\cost^\sigma (\sol_{-i})}{1+\varepsilon} \right) \leq f_i(c,d^\sigma) \leq f_i(c,\cost^\sigma (\sol_{-i})).$$ But notice that 
$$f_i\left(c,\frac{\cost^\sigma (\sol_{-i})}{1+\varepsilon}\right)
\geq \frac{f_i(c,\cost^\sigma (\sol_{-i}))}{1+\varepsilon}.$$ So we have that $$\frac{f_i(c,\cost^\sigma (\sol_{-i}))}{1+\varepsilon} \leq f_i(c,d^\sigma) \leq f_i(c,\cost^\sigma (\sol_{-i})),$$ and therefore $$\frac{1}{f_i(c,d^\sigma)} \in \left[\frac{1}{f_i(c,\cost^\sigma (\sol_{-i}))},  \frac{(1+\varepsilon)}{f_i(c,\cost^\sigma (\sol_{-i}))}\right].$$  So evaluating  expression (\ref{ptaseqn1}) 
using entries from the dynamic programming table instead of the true costs and probabilities incurs a multiplicative error of at most $(1+\varepsilon)$.
\end{proof}

If the input distributions $p_i$ are represented implicitly, such that $\sum_i |C_i|$ is not polynomial in the input size, then we first convert them into distributions with polynomial number of points by combining the probability mass between successive powers of $(1+\varepsilon')$ into buckets (rounding down).  Then we run the above algorithm for discrete distributions so as to obtain a $(1+\varepsilon')$-approximation for the rounded instance. This will be a $(1+\varepsilon')^2$-approximation for the original instance, so if $\varepsilon'$ is chosen such that $(1+\varepsilon')^2 \leq (1+\varepsilon)$, we obtain the desired $(1+\varepsilon)$-approximation.

\subsection{Proof of Theorem \ref{nphardthm}}\label{nphard}
We show that finding the optimal solution, either integer or fractional, to the SBO problem in the scenario model is NP-hard.

The reduction is from \textsc{clique}.  We are given an instance of the \textsc{clique} problem with a graph $G$ containing $n$ nodes and $m$ edges, and a desired clique size $k$. We use $G$ and $k$ to construct an instance $I$ of SBO problem and a number $V$ such that there is a solution to $I$ with expected value of at least $V$ if and only if $G$ contains a clique of size $k$. 

To specify $I$, let us construct a new bipartite graph $H=(L \cup R, E')$ whose right side $R$ contains $n$ nodes corresponding to nodes of $G$, and whose left side $L$ contains $m$ nodes corresponding to edges of $G$. A node in $L$ corresponding to an edge $(u,v)$ is connected to nodes in $R$ that correspond to its endpoints $u$ and $v$.  We first describe the idea of the construction, and later show how to set the parameters to make it work. There will be three parameters, small positive values $\varepsilon$ and $\delta$, and a large value $t$.

All nodes of $H$ are keywords, expensive ones on the left, with $\cpc_i = 1$ for $i\in L$, and cheap ones on the right, with $\cpc_i = \varepsilon$ for $i\in R$. The budget is $K = {k\choose 2}$. The goal will be to force a solution to select $K$ nodes from $L$ that are incident to at most $k$ nodes in $R$, which corresponds to finding a set of $k$ nodes with $K$ edges in $G$, i.e. a $k$-clique.  The scenarios in the SBO problem are as follows. There is a high-probability scenario $\sigma_0$ in which one click comes for each  keyword in $L$. This scenario is sufficiently likely (occurs with probability $1-\delta$) that any integer solution to SBO has to bid for at least $K$ of these keywords. Notice that since $K$ is the budget, it does not make sense to bid on any more than $K$ keywords.  In addition to $\sigma_0$, there are $n$ scenarios $\sigma_1 \ldots \sigma_n$, each occurring with probability $\delta/n$. Scenario $\sigma_i$ contains $K/\varepsilon$ clicks for keyword $i\in R$ and a large number $t$ of clicks for each of $i$'s neighbors from $L$.  

We now explain the intuition for why there is a good integrally-bidding solution for our SBO instance if and only if the graph $G$ contains a $k$-clique. By the way we constructed the low-probability scenarios, if a solution does not bid on any neighbors of $i\in R$, then in scenario $\sigma_i$ it would spend its whole budget on $K/\varepsilon$ cheap clicks at cost $\varepsilon$ each, thus obtaining many clicks.  However, if it bids on any neighbors of $i$, then most of the budget will be spent on the expensive clicks from $L$, resulting in few clicks overall. So bidding on a keyword $l$ from $L$ effectively ruins the scenarios containing $l$'s neighbors in $R$. Recall that the high probability of scenario $\sigma_0$ forces any good integral solution to bid on exactly $K$ keywords from $L$.  As a result, if $G$ contains a $k$-clique, then it is possible to select $K$ keywords corresponding to edges of $G$ that ruin only $k$ scenarios corresponding to nodes of $G$.  However, if there is no $k$-clique, then bidding on any $K$ keywords on the left ruins at least $k+1$ scenarios, thus producing a solution with a lower value.

We now show how to set the parameters of the construction and prove that the reduction works even if fractional bidding on keywords is allowed. First, assume that $G$ contains a $k$-clique.  Then a solution $\sol$ to $I$, with $b_i=1$ for all $i\in R$ and $b_i=1$ for the 
$K$ keywords in $L$ that correspond to edges of the clique, achieves the expected value of at least
$$V = (1-\delta) \cdot K + \frac{\delta (n-k)}{n} \cdot \frac{K}{\varepsilon},$$
where the first term is the value from scenario $\sigma_0$, and the second term is the value from scenarios $\sigma_i$ such that node $i$ in $G$ is not in the clique. Such scenarios are unaffected by the selected nodes in $L$ and therefore get $K/\varepsilon$ clicks each. There is additional value from scenarios $\sigma_i$ for $i$ in the clique, but we disregard it for this lower bound. Thus we get the following claim.

\begin{claim}\label{valifclique}
If $G$ contains a $k$-clique, then there exists a solution $\sol$ to $I$ such that $E[\val(\sol)] \geq V$.
\end{claim}

To ensure that if there is no $k$-clique in $G$, then value $V$ cannot be achieved by any bids, we set the parameters as follows.

\begin{enumerate}
\item Select $\varepsilon$ such that $0 < \varepsilon < \frac{1}{k+1}$.
\item Select $\delta > 0$ small enough that
$\frac{1-\delta}{2}  - \frac{k \delta K}{n \varepsilon} > 0.$
This is possible because the limit of the expression on the left as $\delta \rightarrow 0$ is $\frac{1}{2}$.
\item Let $\alpha = \frac{1}{2m}$. 
\item Choose $t$ large enough that 
$(k+1) \frac{K/\varepsilon + \alpha t}{K+\alpha t} < \frac{1}{\varepsilon}.$ This is possible because $\lim_{t\rightarrow \infty} (k+1) \frac{K/\varepsilon + \alpha t}{K+\alpha t} = k+1 < \frac{1}{\varepsilon}$ by the choice of $\varepsilon$.
\end{enumerate}

\begin{claim}
If $E[\val(\sol)] \geq V$ for some fractional solution $\sol$ to the constructed instance $I$, then there must be at least $K$ keywords $i\in L$ such that $b_i \geq \alpha$.
\end{claim}
\begin{proof}
 Notice that $\sum_{i\in L} b_i \geq (K-1+m\alpha)$ implies that $|\{i\in L~|~b_i \geq \alpha\}|\geq K$, because $b_i$'s are always at most 1. 

What remains to show is that if 
$\sum_{i\in L} b_i < (K-1+m\alpha)$, then $E[\val(\sol)] < V$.
%
%
This follows from the way we defined the parameters. Notice that $\delta\frac{K}{\varepsilon}$ is an upper bound on the value that any solution can obtain from scenarios $\sigma_1 \ldots \sigma_n$. Then 
\begin{eqnarray*}
E[\val(\sol)] < (1-\delta)(K-1+m\alpha) +
\frac{\delta K}{\varepsilon} 
= (1-\delta)(K-\frac{1}{2})+ \frac{\delta K}{\varepsilon} ~=~ V - \frac{1-\delta}{2} + \frac{k}{n} \frac{\delta K}{\varepsilon} ~<~ V,
\end{eqnarray*}
\noindent where the first equality comes from the definition of $\alpha$, and the last inequality from the choice of $\delta$.
\end{proof}

\begin{claim}
Let $X = \{i\in L~|~b_i \geq \alpha\}$. If $H$ contains at least $(k+1)$ nodes in $R$ that have neighbors in $X$, then $E[\val(\sol)]<V$.
\end{claim}
\begin{proof}
For a node $i\in R$, let $\alpha_i = \sum_{j\in N_i} b_j$, where $N_i \subseteq L$ is the set of neighbors of $i$.  Assuming that there are at least $(k+1)$ nodes $i\in R$ such that $\alpha_i \geq \alpha$, we show that $E[\val(\sol)] < V$. Notice that the value of a solution is always maximized by bidding on all keywords in $R$, because that maximizes the number of cheap clicks. So without loss of generality, we assume that $b_i=1$ for all $i\in R$.

In a given scenario $\sigma_i$, $\sol$ gets $\cl^{\sigma_i} = \frac{K}{\varepsilon} + \alpha_i t$, where $K/\varepsilon$ clicks come from keyword $i$, and $\alpha_i t$ come from its neighbors in $L$.  The cost paid in this scenario is $\cost^{\sigma_i} = K+\alpha_i t$, where $K$ is spent for the cheap keywords and $\alpha_i t$ is spent on keywords of cost 1.  Using $(1-\delta)K$ as an upper bound on the value obtained from $\sigma_0$, and remembering that the budget is $K$, we have

\begin{eqnarray*}
E[\val(\sol)] &\leq& (1-\delta)K + \frac{\delta}{n} \sum_i \frac{\cl^{\sigma_i}}{\cost^{\sigma_i}/K} \\
&~=~& (1-\delta)K + \frac{K \delta}{n} \sum_i \frac{K/\varepsilon + \alpha_i t}{K+\alpha_i t} \\
&~\leq~& (1-\delta)K + 
 \frac{K \delta}{n} \left[ (k+1)\cdot \frac{K/\varepsilon + \alpha t}{K+\alpha t} + (n-k-1)\cdot \frac{1}{\varepsilon} \right] \\
& < &  V,
\end{eqnarray*}
where we use the fact that the fraction in the sum increases with decreasing $\alpha_i$, bound $\alpha_i$ by $\alpha$ for $(k+1)$ of the terms and by 0 for the others, and use the choice of $t$ for the final inequality.
\end{proof}

Clearly, if there is no $k$-clique in $G$, then every $K$ edges in $G$ will be incident on at least $k+1$ nodes. So from the preceding two claims, we may conclude that if $G$ does not contain a $k$-clique, then no solution to $I$ has expected value of $V$ or more. Together with Claim \ref{valifclique}, this proves that SBO problem in the scenario model is NP-hard.


\subsection{Proof of Theorem \ref{badprefix}}\label{sec:gappf}
We give an example in which the ratio between the value of the optimal solution and the value of any prefix solution can be arbitrarily large. The example contains $n$ scenarios and $2n$ keywords, numbered 1 through $2n$. The cost per click of keywords increases exponentially, with $\cpc_i = c^i$, for some constant $c>1$.  There is a budget $B>0$. Say that the $n$ scenarios are numbered $\sigma = 1$ to $n$. In scenario $\sigma$, only keywords $2\sigma-1$ and $2\sigma$ receive clicks, and they receive $B/c^{2\sigma-1}$ clicks each. The probabilities of scenarios increase exponentially, and they are equal to $\alpha c^{2\sigma-1}$ for scenario $\sigma$ ($\alpha$ is chosen to make the probabilities sum to 1). The idea here is that in each scenario, there are two types of clicks, cheap and expensive (clicks for the even-numbered keywords are $c$ times more expensive than for their preceding odd-numbered keywords), and there are enough cheap clicks to spend the whole budget. So for a particular scenario $\sigma$, the best thing to do is to bid only on the cheap keyword $2\sigma -1$, which gets $B/c^{2\sigma-1}$ clicks.  Bidding on both keywords exceeds the budget and decreases the number of clicks to $\frac{2}{c+1} (B/c^{2\sigma-1})$. Since the sets of keywords that receive clicks in different scenarios are disjoint, the optimal solution overall (which happens to be integer) is to bid on all the odd-numbered keywords, but not the even-numbered ones.  This gets the maximum number of clicks for each scenario individually, and therefore gets the maximum number of clicks in expectation. The expected number of clicks for the optimal solution is therefore
$$\sum_\sigma \alpha c^{2\sigma-1} \cdot \frac{B}{c^{2\sigma-1}} = n\alpha B.$$

Now consider some prefix solution for this example, either integer or fractional, and the keyword $i^*$ such that $b_i=1$ for $i<i^*$ and $b_i=0$ for $i>i^*$. Let $\sigma^* = \lceil i^*/2 \rceil$ be the scenario containing clicks for keyword $i^*$. Intuitively, the prefix solution ruins the scenarios numbered less than $\sigma^*$ because it bids for both keywords in them, and it ruins the scenarios numbered greater than $\sigma^*$ because it does not bid at all for the keywords in them. As a result, a prefix solution can do well in at most one scenario. It gets the small number of clicks, $\frac{2}{c+1} (B/c^{2\sigma-1})$, for scenarios $\sigma<\sigma^*$, and it gets 0 clicks for scenarios $\sigma>\sigma^*$. It may get up to $B/c^{2\sigma^*-1}$ clicks for $\sigma^*$. So the value of a prefix solution is at most 
\begin{eqnarray*}
\sum_{\sigma<\sigma^*} \alpha c^{2\sigma-1} \cdot \frac{2}{(c+1)} \frac{B}{c^{2\sigma-1}} + \alpha c^{2\sigma^*-1} \cdot \frac{B}{c^{2\sigma^*-1}} 
~\leq~  n \alpha B \left(\frac{2}{c+1} + \frac{1}{n} \right),
\end{eqnarray*}
which can be arbitrarily far from $OPT = n\alpha B$ as $c$ and $n$ increase.

\subsection{Extension to multiple-slot auctions}\label{sec:multslot}
We now sketch the extension of our results to the case when there are multiple slots,
and in particular, we assume the Generalized Second Price (GSP) auction currently
used by search-related advertising engines. 
When advertising slots are allocated by a second-price auction with multiple slots, the 
bid amount for a keyword determines the position of the corresponding ads, which affects 
the number of clicks obtained for this keyword and the cost per click of these clicks.
When a user clicks on the ad in slot $i$,  
the advertiser at slot $i$ is charged the bid amount of the advertiser at slot
$i+1$.\footnote{There are some details, e.g. the cost is typically a small amount
more than the bid of the advertiser at slot $i+1$.}

Let us first focus on a particular keyword $i$ and an auction in which
it participates. If the auction has $k$ available slots, then there are $k$ 
threshold bid amounts, $bid_1 \geq bid_2 \geq ... \geq bid_k$, such that 
bidding any amount in the interval $[bid_j, bid_{j-1})$ places the ad in slot 
$j$, which has a probability of a click (clickthrough rate) $\ctr^j_i$ and a 
cost per click $\cpc^j_i$. Since we are considering the GSP auction, the cost 
per click is not affected by the exact bid amount, as long as it is in the 
specified interval. Both $\ctr^j_i$ and $\cpc^j_i$ are monotone non-decreasing 
step functions of the bid amount. To better represent the options for bidding 
on keyword $i$, we visualize the $k$ possible pairs of 
$(\cpc^j_i  \ctr^j_i, ~\ctr^j_i)$ values on a ``plot'' called a {\em 
landscape}. Notice that when both the axes are scaled by the number of 
queries, then it becomes a plot of $\cl_i$ vs. $\cost_i$, with points for 
different options of how to bid (a more detailed description of landscapes 
appears in~\cite{feldman:budgetoptimization}). Landscapes for multiple auctions for the same 
keyword can be combined to obtain an {\em aggregate} landscape.

Some of our results extend to the model with such aggregate landscapes. 
Roughly speaking, a keyword with a landscape can be viewed as a list of several simple individual keywords with an additional constraint that a solution has to bid on some prefix of this list. 
For the fixed and proportional models, the optimal solutions without landscapes are prefix solutions anyway (by Theorems \ref{fixedthm} and \ref{prefixoptprop}), so if we solve the problem as in the one-slot case, the solution will automatically satisfy the prefix constraint for keywords with landscapes, which means that it will also be optimal for the multiple-slot problem.
In the independent model, however, the approximation ratio of 2 for the prefix solutions (Theorem \ref{2approxthm}), that we prove for the one-slot case, does not extend to the case of  landscapes. This is because some of the ``keywords'' are no longer independent, but are actually the different bidding options for the same keyword. In fact, a prefix solution can be arbitrarily bad compared to the optimal solution, by an example that is very similar to one in Theorem \ref{badprefix}. The only difference is that, instead of keywords $2i-1$ and $2i$ being coupled by occurring in the same scenario, they are coupled by representing the landscape of the same keyword. 
The negative results (Theorems \ref{nphardthm} and \ref{badprefix}) about the scenario model of course still hold for the more general case of multiple-slot auctions.

\end{document}